\documentclass[useAMS,usenatbib,usegraphicx]{mn2e}
\usepackage{times}
\usepackage{epstopdf}

\newif\ifAMStwofonts
\AMStwofontstrue

\newcommand{\simlt}{\lower.5ex\hbox{$\; \buildrel < \over \sim \;$}}
\newcommand{\simgt}{\lower.5ex\hbox{$\; \buildrel > \over \sim \;$}}
\newcommand{\be}{\begin{equation}}
\newcommand{\ba}{\begin{eqnarray}}
\newcommand{\ee}{\end{equation}}
\newcommand{\ea}{\end{eqnarray}}
\newcommand{\Msun}{M$_\odot$}

\title[Constraining the IMF with strong lensing]
{Constraining the low-mass end of the Initial Mass Function with
  Gravitational Lensing}
\author[I. Ferreras et al.]
{Ignacio Ferreras$^1$\thanks{E-mail: ferreras@star.ucl.ac.uk},
Prasenjit Saha$^2$, Dominik Leier$^3$, \and 
Fr\'ed\'eric Courbin$^4$ and Emilio E. Falco$^5$\\
$^1$ Mullard Space Science Laboratory, University College London, 
Holmbury St Mary, Dorking, Surrey RH5 6NT\\
$^2$ Institute for Theoretical Physics, Universty of Z\"urich, 
Winterthurerstr. 190, CH-8057 Z\"urich, Switzerland\\
$^3$ Astronomisches Rechen-Institut, Zentrum f\"ur Astronomie, 
Universit\"at Heidelberg, M\"onchhofsr. 12-14, D-69120 Heidelberg, Germany\\
$^4$ Laboratoire d'astrophysique, \'Ecole Polytechnique F\'ed\'erale 
de Lausanne (EPFL), Observatoire de Sauverny, 1290 Versoix, Switzerland\\
$^5$ Harvard-Smithsonian Center for Astrophysics, 60 Garden Street, 
Cambridge, MA 02138, USA\\
}

\begin{document}
\date{MNRAS Letters, Accepted 2010 August 24.  Received 2010 August
  23; in original form 2010 July 26}
\pagerange{\pageref{firstpage}--\pageref{lastpage}} \pubyear{2010}
\maketitle
\label{firstpage}

\begin{abstract}
  The low-mass end of the stellar Initial Mass Function (IMF) is
  constrained by focusing on the baryon-dominated central regions of
  strong lensing galaxies. We study in this letter the Einstein Cross
  (Q2237+0305), a z=0.04 barred galaxy whose bulge acts as lens on a
  background quasar. The positions of the four quasar images constrain
  the surface mass density on the lens plane, whereas the surface
  brightness (H-band NICMOS/HST imaging) along with deep spectroscopy
  of the lens (VLT/FORS1) allow us to constrain the stellar mass
  content, for a range of IMFs. We find that a classical single power
  law (Salpeter IMF) predicts more stellar mass than the observed
  lensing estimates.  This result is confirmed at the 99\% confidence
  level, and is robust to systematic effects due to the choice of
  population synthesis models, the presence of dust, or the complex
  disk/bulge population mix. Our non-parametric methodology is more
  robust than kinematic estimates, as we do not need to make any
  assumptions about the dynamical state of the galaxy or its
  decomposition into bulge and disk.  Over a range of low-mass power
  law slopes (with Salpeter being $\Gamma=+1.35$) we find that at a
  90\% confidence level, slopes with $\Gamma>0$ are ruled out.
\end{abstract}


%%%%%%%%%%%%%%%%%%%%%%%%%%%%%%%%%%%%%%%%%%%%%%%%
\section{Introduction}

The stellar Initial Mass Function (IMF) is defined as the starting
mass distribution of a stellar population. A large number of studies
have targeted its shape and possible universality \citep[see
e.g.][]{kro02,lar06,kro07,bas10}. It has important implications in the
formation and evolution of galaxies as it affects the chemical
enrichment process and determines the stellar mass content of
galaxies. The seminal work of \citet{salp55} assumed a single power
law distribution between the upper ($\sim$100-120\Msun) and lower mass
cutoffs (0.07\Msun). More recently, the ``local IMF'' has been
constrained in more detail, with a turnover at lower masses
\citep{and08}, and a characteristic mass around
$\sim$1\Msun\citep{lar98}.

Constraining the IMF outside our Galaxy is a challenging task. Data
from unresolved stellar populations are very degenerate with respect
to their properties such as age, metallicity or dust
content. Furthermore, changes in the low-mass end of the IMF do not
translate into significant changes of the photo-spectroscopic
observables, mainly because of the small contribution from low-mass
stars to the net luminosity budget of a galaxy. However, changes in
the low-mass end of the IMF result in measurable changes of the total
stellar mass of a given population. Therefore, any attempt at
constraining the low-mass end of the IMF requires an independent
measure of the stellar mass content of the galaxy. \citet{cap06}
compared the kinematic data from a number of SAURON early-type
galaxies with dynamical models, and concluded that a single power law
for the IMF (\`a la Salpeter) was ruled out. Using gravitational
lensing in the inner, baryon-dominated, regions of early-type
galaxies, \citet{fsb08} showed that a Salpeter IMF produced too much
stellar mass compared to the lensing mass within the same region. In
this paper we extend the gravitational lensing analysis and the
modelling work, to give a more quantitative constraint on the low-mass
end of the IMF. We study in this letter a nearby strong lens, the
Einstein Cross \cite[Q2237+0305][]{EC85}, combining high resolution
NIR imaging with HST/NICMOS and a deep spectrum of the lensing galaxy
taken with the VLT, using the FORS1 instrument, to constrain the
properties of the stellar populations.
 
The reason for choosing this lens is the exceptionally low redshift
(z=0.04) of the lensing galaxy.  This has two advantages.  First, it
helps increase the signal-to-noise. Second, and even more important,
the low lens redshift makes a strong lens from the innermost and
densest part of the galaxy.  The mass-to-light ratio is close to unity
in the lensing region (see Figures 1 and 2 below).  In other words,
the lens is dominated by stars, and hence particularly well-suited to
constraining the IMF.  In contrast, typical galaxy lenses contain a
significant dark-matter contribution.

A different approach using lensing galaxies has been recently taken by
\cite{treu10} and \cite{aug10}. Rather than choosing a single lens
with minimal dark matter and maximal signal-to-noise, they choose a
large sample of lenses, and attempt to subtract off the dark matter by
fitting dark-matter profiles based on simulations.  Their results
favour a Salpeter IMF.  However, a caveat in the method \cite[as noted
by][]{treu10} is that if the assumed dark-matter profile is slightly
too shallow, the stellar mass will be systematically overestimated,
and a ``heavy'' IMF derived.  If this is in fact the case, that would
explain the apparent contradiction with the \cite{fsb08} result, which
excludes a Salpeter IMF for a subset of the same galaxies.

%%%%%%%%%%%%%%%%%%%%%%%%%%%%%%%%%%%%%%%%%%%%%%%%
%%%%%%%%%%%%%%%%  Figure 1   %%%%%%%%%%%%%%%%%%%
%%%%%%%%%%%%%%%%%%%%%%%%%%%%%%%%%%%%%%%%%%%%%%%%
\begin{figure}
  \begin{center}
    \includegraphics[width=3.1in]{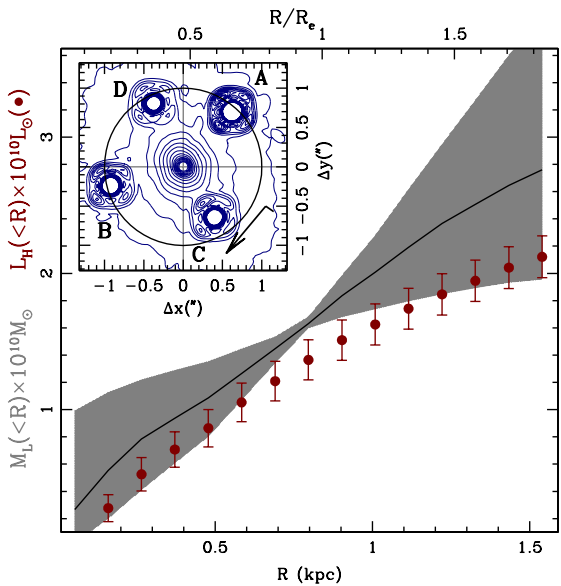}
  \end{center}
  \caption{Cumulative lensing mass profile (shaded region) and
    luminosity (dots). 90\% confidence levels shown. Notice the
    butterfly-shape of the uncertainty for lensing mass, which 
    is minimal at the position on the lens plane of the 
    images of the background source. The inset shows
    a contour map of the F160W HST/NICMOS image. At the lens position
    1 arcsec maps into 0.77~kpc. The arrow points North, the small
    segment from the base of the arrow points East.
   \label{fig:prof}}
 \vskip-0.1in
\end{figure}
%%%%%%%%%%%%%%%%%%%%%%%%%%%%%%%%%%%%%%%%%%%%%%%%

%%%%%%%%%%%%%%%%%%%%%%%%%%%%%%%%%%%%%%%%%%%%%%%%
\section{Constraining the IMF}
In order to constrain the low-mass end of the IMF, we need to combine
stellar population synthesis with an independent estimate of total
mass in a system whose total mass content is preferrably (but not
necessarily) dominated by the stellar mass. We note that even though
baryonic, non-stellar matter (e.g. gas and dust) will contribute to
the mass budget, the constraints on the IMF imposed in this letter go
in the sense that rejected choices of the IMF predict {\sl more}
stellar mass than the total mass determined by the lens geometry.  We
use strong gravitational lensing as the independent estimator of
mass. Kinematics-based methods have been used recently to quantify
the total mass in the inner regions of galaxies \citep[see
e.g.][]{cap06,coc09}. However, such methods rely on a particular
definition of the dynamical system.

The lens is modelled using the {\em PixeLens\/} method
\citep{sw04,coles08}, whereby the projected
% and may be summarized as follows.  The projected
mass distribution is reconstructed as a pixellated mass map, which
%, which reproduces the image positions exactly.  
%The mass map 
is not limited to any parametric form \citep[cf.][]{trott10} but can
be any non-negative distribution subject to the following prior
conditions.  (i)~The mass map is symmetric under a $180^\circ$
rotation about the centre of brightness.  (ii)~The local density
gradient must point no more than $45^\circ$ away from the centre,
which is assumed to coincide with the light peak. (iii)~No pixel,
except the central one, is allowed to be more than twice the mean of
its neighbours.  (iv)~The circular average of the projected density,
falls off as $R^{-1/2}$ or steeper.  In other words, the effective
three-dimensional mass profile should be $\propto r^{-3/2}$ or
steeper.

The above requirements do not give a unique mass map, because
solutions of the lens equation are highly non-unique
\citep{fal85,pras00,lie08}.  Accordingly, the possible ``model space''
is sampled by an ensemble of 300 models.  A useful property of the
ensemble, due to the linear nature of the observational and prior
constraints, is that the ensemble average is also a valid model; for
comparisons of ensemble averages with individual models in {\em
  PixeLens,} see Figures~7 and 8 in \cite{sw06}.
%%%%%%

%%%%%%%%%%%%%%%%%%%%%%%%%%%%%%%%%%%%%%%%%%%%%%%%
%%%%%%%%%%%%%%%%  Figure 2   %%%%%%%%%%%%%%%%%%%
%%%%%%%%%%%%%%%%%%%%%%%%%%%%%%%%%%%%%%%%%%%%%%%%
\begin{figure}
  \begin{center}
    \includegraphics[width=3.1in]{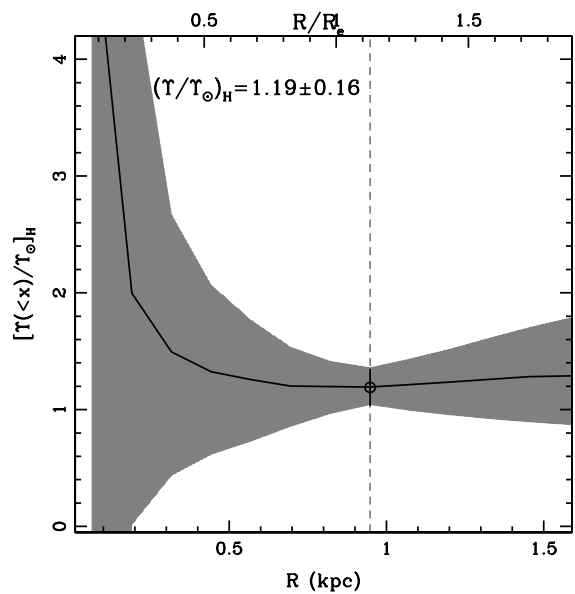}
  \end{center}
  \caption{Predicted H-band mass to light ratio combining the analysis of the 
    lensing data and the luminosity profile of the lens. The dot corresponds 
    to the value chosen in this paper to constrain the stellar M/L, with 
    mean and standard deviation as labelled.
    \label{fig:profML}}
\vskip-0.1in
\end{figure}
%%%%%%%%%%%%%%%%%%%%%%%%%%%%%%%%%%%%%%%%%%%%%%%%

For this study, we select the Einstein Cross \citep{EC85}, where the
bulge of a barred Sab galaxy at z=0.0394 acts as a lens, generating a
quadruple image of a background z=1.695 QSO. Our reference image is
taken from the CASTLES survey\footnote{\tt http://www.cfa.harvard.edu/castles}.  
It is a 384 second exposure with NIC2/HST through passband F160W.  The
lensing analysis only uses the redshifts of the QSO and of the lens,
along with the QSO image positions to constrain the surface mass
density on the lens plane. The NICMOS image is also used to
reconstruct the surface brightness profile of the lens. We use GALFIT
\citep{galfit} in order to remove the unresolved images of the
background QSO \citep[see][for details]{lfsf10}. Figure~\ref{fig:prof}
shows the cumulative mass and H-band luminosity profiles of the
lens. The uncertainties in lensing mass and luminosity are given by
the shaded region and errorbars, respectively. Notice the lens
geometry allows us to constrain in detail the cumulative mass within
$R\simlt 0.75$~kpc, roughly around one effective radius (top axis).

%%%%%%%%%%%%%%%%%%%%%%%%%%%%%%%%%%%%%%%%%%%%%%%%
%%%%%%%%%%%%%%%%  Figure 3   %%%%%%%%%%%%%%%%%%%
%%%%%%%%%%%%%%%%%%%%%%%%%%%%%%%%%%%%%%%%%%%%%%%%
\begin{figure}
  \begin{center}
    \includegraphics[width=3.1in]{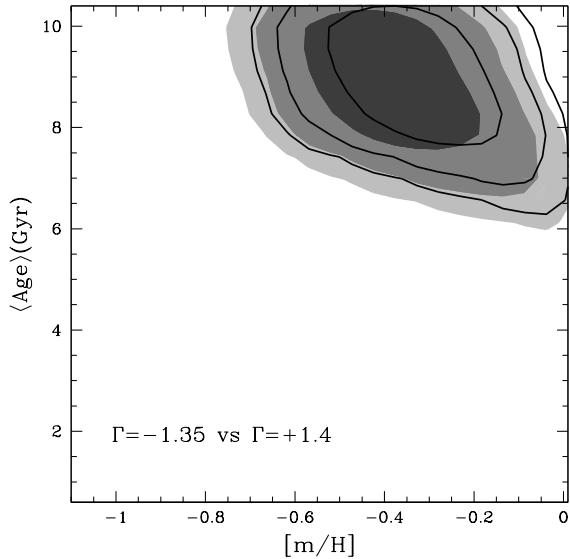}
  \end{center}
  \caption{The 68, 95 and 99\% confidence levels in metallicity
    and average age are shown for the $\tau$ models described in the text,
    using the P\'egas\'e-HR (High Resolution) library.
    The greyscale corresponds to an IMF with slope
    $\Gamma=+1.35$ (i.e. Salpeter), whereas the lines represent the
    other extreme explored in this paper ($\Gamma=-1.4$).
   \label{fig:chi1}}
\vskip-0.1in
\end{figure}
%%%%%%%%%%%%%%%%%%%%%%%%%%%%%%%%%%%%%%%%%%%%%%%%

The range of lens models has a characteristic butterfly shape, which
can be understood as follows.  For a pure Einstein ring, the enclosed
mass $M(<R)$ is perfectly constrained at the Einstein ring, but
degenerate at any other radius.  Now, the prior allows mass profiles
ranging from $R^{-1/2}$ to a point mass, hence $M(<R)$ can range from
$R^{3/2}$ to flat.  The maximal and minimal values of steepness,
combined with the Einstein-radius constraint, shape the allowed range
of models. In practice, because real lenses are not pure Einstein
rings, the range of models is not quite so extreme, but retains the
butterfly shape.
%%%%%% 
The inset shows the H-band frame of the lens and quasar images along
with their standard identifications. The mass and luminosity profiles
are compared to obtain the H-band mass-to-light ratio ($\Upsilon_{\rm
  H}$), as shown in figure~\ref{fig:profML}, including the uncertainty
as a shaded region. The point where the errorbar is smallest is chosen
to constrain the low-mass end of the IMF. The figure shows that within
this region, the lensing galaxy has an average $\Upsilon_{\rm
  H}=1.19\pm 0.16$ (90\% confidence level).

Our next step involves the construction of a large number of composite
stellar populations, following an exponentially decaying star formation
history. Hence, each model is characterised by the
formation epoch, decay timescale and metallicity. We also need to
consider different sets of models for a number of choices of the IMF.
Notice these are composite models and not simple stellar populations
(SSPs), therefore giving more robust mass-weighted
values. SSP-determined values are easily affected by the presence of
recent star formation, even with small amounts of young stars,
resulting in -- sometimes significantly -- lower M/L ratios \citep[see
e.g. ][]{bmc10}.

%%%%%%%%%%%%%%%%%%%%%%%%%%%%%%%%%%%%%%%%%%%%%%%%
%%%%%%%%%%%%%%%%  Figure 4   %%%%%%%%%%%%%%%%%%%
%%%%%%%%%%%%%%%%%%%%%%%%%%%%%%%%%%%%%%%%%%%%%%%%
\begin{figure}
  \begin{center}
    \includegraphics[width=3.1in]{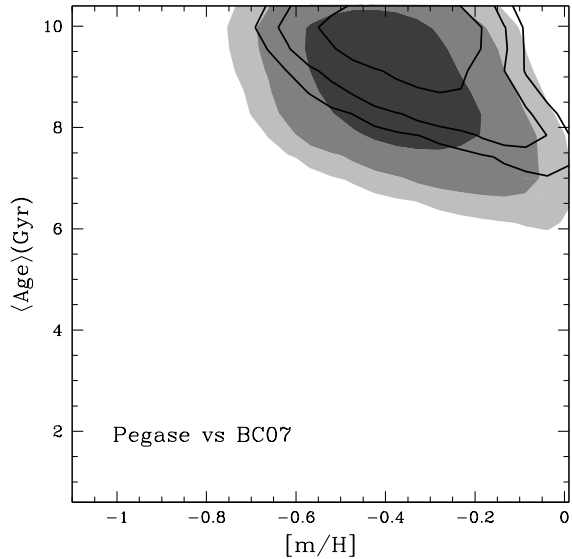}
  \end{center}
  \caption{Same as figure~\ref{fig:chi1} comparing
    two different sets of models, both assuming a 
    Salpeter IMF. The greyscale corresponds to 
    P\'egas\'e-HR, and the lines are the predictions
    of the BC07 models (see text for details).
   \label{fig:chi2}}
\vskip-0.1in
\end{figure}
%%%%%%%%%%%%%%%%%%%%%%%%%%%%%%%%%%%%%%%%%%%%%%%%

In previous work \citep{fsw05,fsb08,lfsf10} we used broadband
photometry to constrain the mass-to-light ratio of the stellar
populations. We showed that acceptable uncertainties can be obtained
on M/L even though one should not use broadband photometry alone to
constrain the age and metallicity of the unresolved populations
\citep[see e.g.][]{gala09}. However, in order to minimise the
uncertainties in the estimates of M/L, we use deep spectroscopic
observations of the Einstein Cross from VLT/FORS1 with grism G300V,
giving a spectral resolution of R$\sim 400$ at 5900\AA\
\citep{vlt}.
The spectra are taken at two slit positions: mask 1 (t$_{\rm EXP}$=11.70h) 
at PA$=-56.5^\circ$, passing through images A and D;
and mask 2 (t$_{\rm EXP}$=11.25h) at PA$=-68.5^\circ$, passing through
images B and C (see figure~\ref{fig:prof}).
The 2D spectra are separated into a point source -- the quasar
spectrum -- and an extended source -- the lens. The method leaves no
trace of the emission line spectrum of the quasar in the lens
spectrum, which can be seen in figure~7 of \citet{vlt}.

We measure the equivalent widths (EWs) of a number of age and
metallicity sensitive spectral features: Balmer H$\beta$, H$\gamma$,
H$\delta$, Mgb, Fe5270, Fe5335, G4300 and the 4000\AA\ break. The
EWs are determined using a recent definition of the continuum
\citep{bmc10} that reduces the contamination in EW estimates from
neighbouring lines, improving on the discrimination between age and
metallicity.

We also have at our disposal the HST/NICMOS (F160W) and HST/WFPC2
(F555W and F814W) images. Note we already use the F160W image to
determine the maximum $\Upsilon_H$ from stars. However, we can combine
these three bands to generate two colours.  At the effective radius,
the colours are $V-I=1.28$ and $I-H=1.89$ (Using Vega as zero-colour
reference). The 1$\sigma$ photometric uncertainties stay around
0.05~mag in both colours. The colour gradient is very shallow: over
the R$<2$R$_e$ range, the gradients $\Delta$Colour$/\Delta (R/R_e)$
are $-0.003$ and $0.05$ in $V-I$ and $I-H$, respectively. The flat
colour profile confirms the homogeneity of the stellar populations in
this lens.  The colours help further constrain the stellar
populations, although we emphasize that the inclusion of the colours
does not improve significantly the estimates based on EWs, as
expected. 

From the available stellar population synthesis models, we chose the
high spectral resolution version of P\'egas\'e \citep{Peg}. These models allow
the user to recalculate the spectra of simple stellar populations for
an arbitrary choice of IMF. Our goal is to constrain the shape of the
low-mass end of the IMF. Hence, we decided to model the IMF as two
power laws, defined as follows:
\begin{equation}
\frac{d{\rm N}}{d\log {\rm M}} \propto \left\{
     \begin{array}{ll}
                {\rm M}^{-\Gamma} &  0.1<{\rm M}/{\rm M}_\odot<1\\
                & \\
                {\rm M}^{-1.35} &  1<{\rm M}/{\rm M}_\odot<100
              \end{array}
       \right. 
\label{eq:imf}
\end{equation}
\noindent
The above definition follows \citet{kro07}, although we select the
Salpeter IMF slope \citep{salp55} at the high mass end and
choose 1\Msun\  for the position of the knee.  For masses below this
knee, we parameterise the IMF as a power law with index $\Gamma$. We
explore a wide range of values, from $\Gamma=-1.4$ to $+1.4$ (where a
Salpeter IMF corresponds to $\Gamma=+1.35$). For each choice of
$\Gamma$, we generate a set of base models (simple populations) that
are combined into composite $\tau$-models as described above, covering
a grid of $48\times 48\times 48$ models over a range of formation
epoch ($2\leq z_{\rm FOR}\leq 10$); decay timescale
($-1\leq\log(\tau/1{\rm Gyr})\leq +1$) and metallicity ($-1\leq
[m/H]\leq +0.3$). Hence, our methodology takes into account a spread
in the age distribution, an important issue when dealing with the
stellar populations of galaxies \citep[see e.g.][]{bmc10}.  Each
choice of star formation history gives a set of EWs and colours that
are compared with the observations via a standard $\chi^2$-defined
likelihood to infer the probability distribution function. The
synthetic spectra are broadened to the spectral resolution of the
observations, taking also into account the observed velocity
dispersion of the lens, $\sigma=166\pm 2$~km/s \citep{vdv08}.  Also
notice that the constraint on the M/L ratio shown in
figure~\ref{fig:profML} corresponds to the {\sl observed frame}
H-band. The models take into account the minor K-correction needed,
given the redshift of the lens.

%%%%%%%%%%%%%%%%%%%%%%%%%%%%%%%%%%%%%%%%%%%
\section{The systematics of stellar mas estimates}

It is important to understand the systematic effects caused by the
choice of population synthesis models in the estimates of stellar
mass. Firstly, we consider the effect that a different choice of slope
would have on the predicted average age and metallicity of the lens
galaxy.  Figure~\ref{fig:chi1} shows the probability distribution of
the average age and metallicity between the two extreme choices of
IMF: namely a Salpeter IMF ($\Gamma=+1.35$, greyscale) and a strong
suppression of low-mass stars ($\Gamma=-1.4$, lines). One can see that
within error bars, the average age and metallicity are unchanged,
corresponding to an old ($\simgt 8$~Gyr) population. 

Different population synthesis models rely on their choices of stellar
evolution prescriptions, isochrones, and stellar libraries.  We
quantify this effect in the following two
figures. Figure~\ref{fig:chi2} shows the age and metallicity
probability distribution for a Salpeter IMF using two independent
models: P\'egas\'e (grayscale) and Bruzual \& Charlot (2003;
lines). The latter tends to give slightly older ages, although the
confidence levels overlap, eliminating the chance of a systematic
offset caused by a different age or metallicity.

\begin{figure}
  \begin{center}
    \includegraphics[width=3.1in]{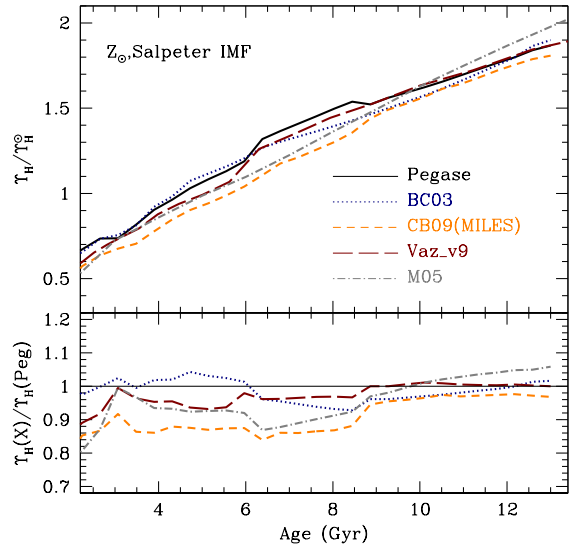}
  \end{center}
  \caption{Comparison of the H-band stellar mass-to-light ratio
    predicted by a number of population synthesis models. For ease of
    comparison, we consider here only simple stellar populations with
    solar metallicity and a Salpeter IMF. The models are P\'egas\'e
    \citep[black solid line,][]{Peg}; two different flavours of the
    GALAXEV models: BC03 \citep[blue dotted line][]{BC03}; and the recent,
    unpublished 2009 models including the MILES spectral library
    \citep[orange short dashed lines, private communication,
    and][]{miles}; Vazdekis/MILES models v.9 \citep[long
    dashed lines][]{vz10} and \citet[grey dot-dashed lines,][]{M05}.
    The bottom panel shows the ratio with respect to the fiducial
    P\'egas\'e-HR models used in this paper.
    \label{fig:popsyn}}
\vskip-0.1in
\end{figure}

To emphasize this point, figure~\ref{fig:popsyn} illustrates the
predictions in $\Upsilon_H$ for a range of Simple Stellar Populations
of solar metallicity -- as suggested by the best fits shown in the
previous figures. Various population synthesis models are shown, as
labelled. Hence, for the age range considered to be a best fit
($\simgt 8$~Gyr), the systematic difference between these models
cannot be larger than $\sim 10$\%. Another possible systematic
source would be the presence of dust, mainly in the disc of the lens
galaxy. \citet{vlt} showed that the $V$-band differential extinction
caused by the lens on the background quasar is in the range 0.1--0.3
mag. Using a standard extinction law \citep[e.g.][]{fitz99}, we estimate
an attenuation no larger than about 0.05 mag in the H band. Hence,
the net systematic uncertainty on the determination of the stellar masses 
is $\sim 15-20$\%.

%%%%%%%%%%%%%%%%%%%%%%%%%%%%%%%%%%%%%%%%%%%%%%%%
%%%%%%%%%%%%%%%%  Figure 5   %%%%%%%%%%%%%%%%%%%
%%%%%%%%%%%%%%%%%%%%%%%%%%%%%%%%%%%%%%%%%%%%%%%%
%%%%%%%%%%%%%%%%%%%%%%%%%%%%%%%%%%%%%%%%%%%%%%%%

%%%%%%%%%%%%%%%%%%%%%%%%%%%%%%%%%%%%%%%%%%%
\section{Results}

\begin{figure}
  \begin{center}
    \includegraphics[width=3.1in]{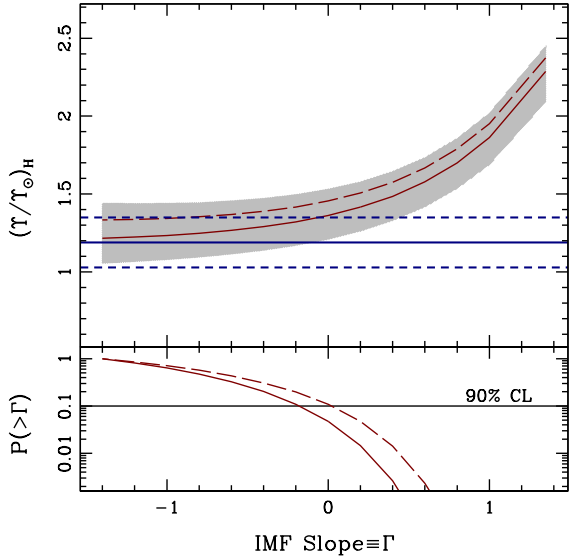}
  \end{center}
  \caption{{\sl Top:} Constraints at the 90\% confidence level from
    the observed equivalent widths of Balmer lines (H$\beta$,
    H$\gamma$, H$\delta$); the metal sensitive [MgFe] index; and the
    $V-I$ and $I-H$ colours. The result is shown as a function of the
    low-mass slope of the IMF. The shaded region corresponds to the
    $\Upsilon_H$ given by the P\'egas\'e-HR models used in this paper.
    The horizontal lines give the 90\% confidence level for the
    predicted M/L using lensing data and the observed photometric
    profile of the lens (see figure~\ref{fig:profML}). The solid (long
    dashed) curved lines correspond to slit 1 and 2, respectively,
    with the shaded region extending over the 90\% confidence level
    for both slits 1 and 2. {\sl Bottom:} Cumulative joint probability
    of the IMF slope, given the lensing estimate of $\Upsilon_H$ and
    the population synthesis constraints. A Salpeter IMF
    ($\Gamma=+1.35$ is ruled out at more than the 99\% confidence
    level. At a 90\% level, slopes $\Gamma>0$ are rejected as well.
    \label{fig:exp}}
\vskip-0.1in
\end{figure}

Figure~\ref{fig:exp} shows the constraint on the stellar $\Upsilon_H$
from the photo-spectroscopic data, as a function of IMF low-mass
slope. For reference, the upper limit to the $\Upsilon_H$ from lensing
is shown as a horizontal line along with the 90\% confidence level
(dashed lines). A Salpeter IMF ($\Gamma=+1.35$) clearly gives a
stellar $\Upsilon_H$ that is too high with respect to the lensing
estimate. The best fits are shown as curved solid and long-dashed
lines for the analysis of spectra from masks 1 and 2, respectively.
The result is compatible within the 90\% confidence level (grey shaded
area). Furthermore, our data can constrain in more detail the low-mass
end of the IMF. The lower panel of figure~\ref{fig:exp} shows the
cumulative probability distribution function, P$(>\Gamma)$. At a 90\%
confidence level, slopes $\Gamma>0$ are ruled out.

Our modelling has taken an ``integrated approach'' to the presence of
disk and bulge. Indeed, both the spectroscopic and the photometric
data have not been separated into a bulge and a disk component. Hence,
the modelling of the data corresponds to an ``average'' stellar
population which would be dominated by the old component, given that
most of the light originates from the central region of the galaxy, a
point confirmed by the old ages and high metallicities predicted by
the modelling (see figures~\ref{fig:chi1} and \ref{fig:chi2}). The
photometric analysis is done in a non-parametric way, hence we compute
stellar masses from this ``composite population'' of disk and bulge
stars in a consistent way.  Were we to remove the presence of the disk
population, the older bulge stars would result into higher
$\Upsilon_H$ therefore making a Salpeter IMF even more incompatible
with the observations. Finally, a significant presence of dark matter
in the inner regions of galaxies \citep[see e.g.][]{tor09} will make
the rejected models even more unlikely, since the analysis presented
here predicts more stellar mass than lensing mass for those rejected
models.  Therefore, our conclusions are robust with respect to the
complex populations of disk and bulge stars, or to the presence of
dark matter.

%%%%%%%%%%%%%%%%%%%%%%%%%%%%%%%%%%%%%%%%%%%%%%%%
%%%%%%%%%%%%%%%%  Figure 6   %%%%%%%%%%%%%%%%%%%
%%%%%%%%%%%%%%%%%%%%%%%%%%%%%%%%%%%%%%%%%%%%%%%%
%%%%%%%%%%%%%%%%%%%%%%%%%%%%%%%%%%%%%%%%%%%%%%%%

%%%%%%%%%%%%%%%%%%%%%%%%%%%%%%%%%%%%%%%%%%%%%%%%
\section*{Acknowledgments}
FC is partially supported by the SNSF (Switzerland).

%%%%%%%%%%%%%%%%%%%%%%%%%%%%%%%%%%%%%%%%%%%%%%%%
%%%%%%%%%%%%%%%   REFERENCES   %%%%%%%%%%%%%%%%%%%%%%
%%%%%%%%%%%%%%%%%%%%%%%%%%%%%%%%%%%%%%%%%%%%%%%%

\label{lastpage}
\end{document}